\documentclass[a4paper,11pt]{article}
\pdfoutput=1
\usepackage{jheppub}
\usepackage{subfig}
\usepackage{comment}
\usepackage{array,setspace,mathrsfs,amsfonts,amsmath,yfonts,dsfont,bbm,colonequals}
\usepackage{graphicx} 
\usepackage{amsmath}
\usepackage{amssymb}
\usepackage{amsfonts}
\usepackage{appendix}
\usepackage{comment}
\usepackage{url}
\usepackage[outline]{contour}
\usepackage{color}
\newcommand\hare[1]{\textcolor{blue}{(hare: #1)}}
\newcommand\sarah[1]{\textcolor{green}{(Sarah: #1)}}

\title{There is more to the de Sitter horizon than just the area}

\author{Willy Fischler*, Hare Krishna* and Sarah Racz$^{*\dagger}$}

\affiliation{
$*$ Weinberg Institute, Department of Physics, University of Texas at Austin, Austin TX 78712, USA\\
$\dagger$ Sarah Lawrence College, Bronxville NY 10708, USA\\}

\emailAdd{fischler@physics.utexas.edu}\emailAdd{hkrishna.phy@gmail.com}\emailAdd{sracz@sarahlawrence.edu}

\abstract{It is well known that the area of the de Sitter cosmological horizon is related to the entropy of the bulk spacetime. Recent work has however shown that the horizon encodes more information about the bulk spacetime than just the entropy. In this work, we show that the horizon contains all of the gauge invariant (diffeomorphism and $U(1)$) information about (static albeit unstable) configurations of charged and rotating objects placed deep inside the de Sitter spacetime. We study highly symmetric objects, such as dipoles and cubes, built of objects with electric charge and angular momentum at their vertices. We show how these configurations affect the geometry of the cosmological horizon and imprint detailed information about the objects in the bulk onto the cosmological horizon.} 
\begin{document}
\hfill       UT-WI-43-2024

\maketitle

\flushbottom
\section{Introduction}
In the long-sought theory of quantum gravity, the holographic principle has emerged as a key concept. For the past three decades, the holographic principle has been the subject of intense efforts, particularly through the AdS/CFT correspondence (see \cite{hep-th/9905111} and references therein). The partial successes of the AdS/CFT approach have yet to be replicated in general space-times, in particular in de Sitter and flat spacetime cases. \\

The de Sitter spacetime is of particular interest as it appears that our universe is undergoing accelerated expansion consistent with an equation of state of a small positive cosmological constant (asymptotically de Sitter spacetime). If we endeavor to understand the quantum nature of gravity in our own universe, we should set our aims for a quantum theory of de Sitter spacetime. Though efforts have been made towards this goal \cite{hep-th/0102077, Fischler2000, Banks2000, hep-th/0007146, hep-th/0609062, hep-th/9806039, 2306.05264,2109.01322,2209.09999}, there remains work in achieving a satisfactory quantum theory of de Sitter spacetime. \\

In the 70's, Gibbons and Hawking realized that de Sitter spacetime has an entropy and temperature \cite{Gibbons1977}. The entropy of the empty de Sitter spacetime is given by the Bekenstein-Hawking area law for the cosmological horizon
\begin{equation}
\label{eq:BekensteinHawking}
S_{dS} = \frac{A_{\mathcal{CH}}}{4 G},
\end{equation}
which is determined by the de Sitter radius $l$.
The Schwarzschild-de Sitter (SdS) solution describes a black hole in \scalebox{1.00}{\textbf{unstable equilibrium}} embedded in the de Sitter spacetime. As compared to empty de Sitter spacetime, the Schwarzschild-de Sitter solution reduces the area and thus the entropy of the cosmological horizon. For black holes sufficiently smaller than the de Sitter radius, $m \ll l$, \ the area of the Schwarzschild-de Sitter cosmological horizon is given by
\begin{equation}
    A_{SdS} = A_{dS} - l m,
\end{equation}
where $m$ is the mass of the black hole within the bulk. The entropy of the Schwarzschild-de Sitter cosmological horizon has a deficit of $l m$ as compared to the empty de Sitter horizon. It has been well understood that empty de Sitter spacetime is maximally entropic and that any object or excitation in the spacetime reduces the entropy. Albeit observer-dependent, the cosmological horizon has emerged as a natural location for the quantum degrees of freedom of de Sitter spacetime \cite{hep-th/0007146,hep-th/0102077,hep-th/0609062,hep-th/0212209, Banks2000,  hep-th/9806039, 2306.05264,2109.01322,2209.09999, Fischler2000}. Various proposals for the quantum theory of de Sitter, such as the holographic space-time (HST) proposal of Banks and Fischler \cite{Banks:2001px,Banks:2003ta,Banks:2015iya} and the double-scaled SYK (DSSYK) proposal of Susskind \cite{2209.09999}, treat the cosmological horizon as the holographic screen of the theory and are formulated as `static patch holography'. In this work, we study deformations of the holographic screen sourced by objects with charge and rotation within the de Sitter bulk. \\

Recently, Fischler and Racz \cite{Fischler:2024cgm} showed that the cosmological horizon responds to and encodes information about extended objects placed within the de Sitter bulk. They considered configurations of masses arranged on the vertices of Platonic solids and found that the cosmological horizon deformed to the dual polyhedron of the bulk-matter configuration. Interestingly enough, the horizon changes by an area-preserving deformation, so the entropy of spacetime only depends on the total mass present within the de Sitter bulk. However, the shape of the cosmological horizon, which can be determined through its discrete symmetry group, even encodes the size of the extended object within the bulk. Thus, all of the information of the objects within the bulk can be determined by measurements of just the cosmological horizon. While the authors primarily focused on Platonic solids, the results are expected to hold for more general configurations of masses in unstable-static equilibrium. While the configurations are in unstable-equilibrium, we remind the reader that the Schwarzschild-de Sitter solution is \textbf { unstable} as well, and nevertheless, important lessons have been drawn from that spacetime \cite{Fischler2000, Banks2000, hep-th/0007146,hep-th/0609062,2109.01322}.

In this article, we extend the Fischler-Racz analysis to include electric charges and rotation. We use the tools of black hole perturbation theory, as formulated by Regge and Wheeler \cite{Regge1957}, applied to the cosmological horizon to find its deformations. While we focus on dipole and cube configurations of matter, we believe our conclusions generalize to more complicated objects. Through our analyses, we confirm that the cosmological horizon is the dual shape of the bulk configuration. The difference here is that the energy stored in the electric field of charged objects contributes to the horizon deformation. We will find that the electric fields between charged objects satisfy the same symmetry group as those objects. 

Additionally, we obtain the gauge invariant (both EM and diffeomorphism) charges on the horizon. Electric field lines originate from the charges placed within the bulk and end at the `induced' charges on the cosmological horizon. The field lines are consistent with Gauss's law in de Sitter spacetime. The induced charge on the horizon ultimately allows for the reconstruction of bulk charge data from the cosmological horizon. If we instead give the objects placed within the bulk angular momentum ($a=J/M$ small), we find corresponding induced rotation on the cosmological horizon.\\

\textbf{Outline:-} In section \ref{configuration equilibrium}, we find the equilibrium position of the charged objects within the de Sitter bulk, which fixes the object length scale $d$ in terms of the other parameters $m,l$, and $q$. We do not calculate different equilibrium positions of rotating objects since the angular momentum coupling of objects is a subleading effect. In section \ref{charged formalism}, we review the Regge-Wheeler formalism which we use to find the metric perturbations associated with charged and rotating objects. In sections \ref{charged objects} and \ref{rotating formalism}, we calculate explicit examples of metric perturbations due to configurations of masses with charge and rotation. Using these perturbations, we find the shape and location of the new cosmological horizon. We confirm that the horizon inherits the same discrete symmetry group as the objects placed within the bulk. We also see that for charged matter configurations, the horizon inherits information about the individual charge of the constituent objects. For rotating objects, we find that the cosmological horizon has information about the angular momentum of the individual objects. We end the article with a discussion of our findings and future directions in section \ref{discussions}. The details about the decomposition of the stress tensor in tensor harmonics are relegated to the appendix \ref{stress_decom}.

\section{Static configurations of charged massive objects in de Sitter spacetime}
\label{configuration equilibrium}
We begin our analyses by finding static configurations of multiple charged masses within the de Sitter bulk. We are careful not to `overcharge' our masses, which would introduce naked singularities in our spacetime. There are three forces at play here: the force due to cosmic expansion, gravitational attraction, and the electromagnetic interaction between any two objects in a configuration. The net gravitational and electric magnetic force between bulk objects must point radially inwards so that they may be canceled by the force of cosmic repulsion. These bulk configurations are in unstable equilibrium. Perturbations can either cause the objects to coalesce into a single object at the center of the static patch or to fly towards the cosmological horizon.

For masses sufficiently smaller than the de Sitter radius in Planck units (which we adopt throughout this paper), the net force acting on each object is well approximated in the Newtonian regime by the Newton-Hooke equation
\begin{eqnarray}
\label{eqn:NewtonHooke}
    m_i \frac{d^2 \vec{x}_i}{dt^2}=m_i  \frac{\vec{x}_i}{l^2}- \sum_{j\neq i}^N\frac{m_i m_j(\vec{x}_i-\vec{x}_j)}{|\vec{x}_i-\vec{x}_j|^3}-\sum_{j\neq i}^N\frac{q_i q_j(\vec{x}_i-\vec{x}_j)}{|\vec{x}_i-\vec{x}_j|^3}=0,
\end{eqnarray}
For objects in equilibrium, the total force acting on each object is zero. Arrangements of masses satisfying (\ref{eqn:NewtonHooke}) are called central configurations in the literature and have been extensively studied in \cite{math-ph/0303071,hep-th/0201101,hep-th/0308200}.

The center of mass of each configuration of masses within the de Sitter bulk needs to lie at $r=0$ otherwise, the entire configuration will fly toward the horizon. For simplicity, we take all the masses and charges to be equal as $m_i=m$ and $|q_i|=q$, and $|q|<<m$. The placement of positive and negative charges on the vertices of polyhedra in the bulk needs to be done carefully if the configuration is to be in equilibrium. We find that the charges must have the same symmetry group as the mass configuration to remain in equilibrium. To recover the length scales in \cite{Fischler:2024cgm}, we simply set the charge $q=0$

We now turn our attention to three cases: that of the dipole, the cube, and a cube superimposed with a mass located at $r=0$, which we refer to as the ``crystalline atom". The results for charged mass configurations can be extended to the Platonic solids studied in \cite{Fischler:2024cgm}.

\subsection*{A dipole:}
We consider a dipole consisting of two masses  $m_i=m,\,\,\vec{x}_i=\frac{d}{2} \hat{z}$ with opposite charges ($\pm q$) at each end. Solving the equilibrium condition given by the equation. \ref{eqn:NewtonHooke} we find the separation of the charged masses to be 
\begin{eqnarray}
\label{dipole equi}
    - \frac{  d}{2 l^2}+  \frac{  m}{  d^2}+ \frac{  q^2}{ m d^2}=0, \quad \implies d =\frac{\sqrt[3]{2} \sqrt[3]{l^2 \left(m^2+q^2\right)}}{\sqrt[3]{m}}.
\end{eqnarray}
\subsection*{A cube:}
\begin{figure}
    \centering
    \includegraphics[width=0.5\linewidth]{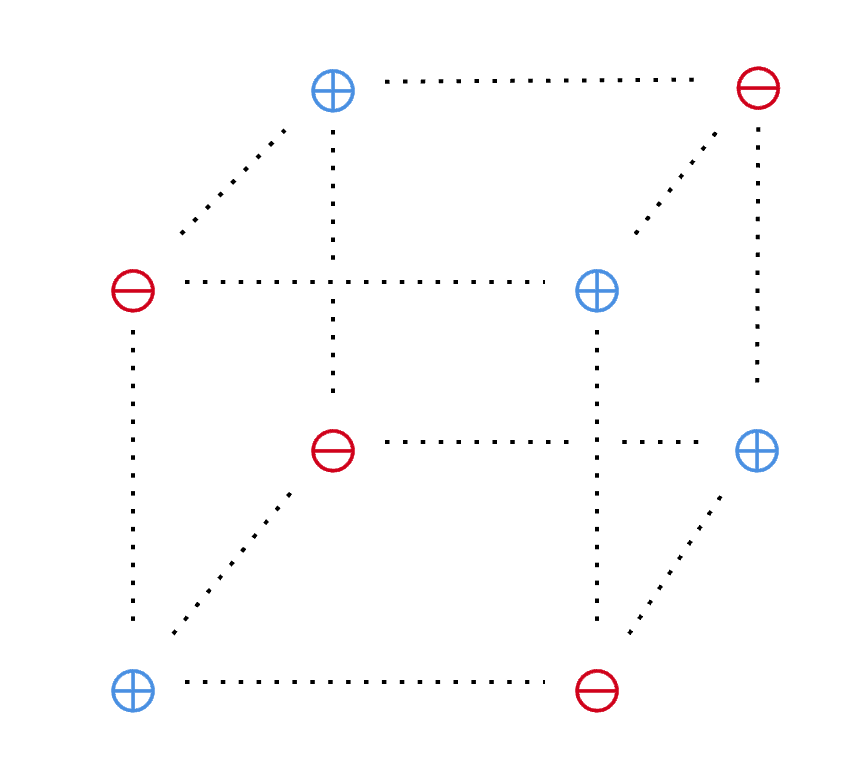}
    \caption{This configuration of charges and masses on the vertices of a cube. The effective Coulomb forces on any of the charges point towards the center of the cube along the main diagonal. The gravitational force points towards the center. The sum of these two forces is balanced by cosmological repulsion. }
    \label{fig:cube config}
\end{figure}

Next, we study the configuration shown in fig.  \eqref{fig:cube config} of masses with charge $\pm q$ placed on the vertices of a cube. We find that the configuration shown of alternating charges is the only configuration where the net force on each particle points toward the center of the static patch and can thus be balanced by the force of cosmic expansion. We note that this alternating charge configuration has the discrete symmetry group as the masses are placed at the cube's vertices. 

\begin{figure}
    \centering
    \includegraphics[width=0.43\linewidth]{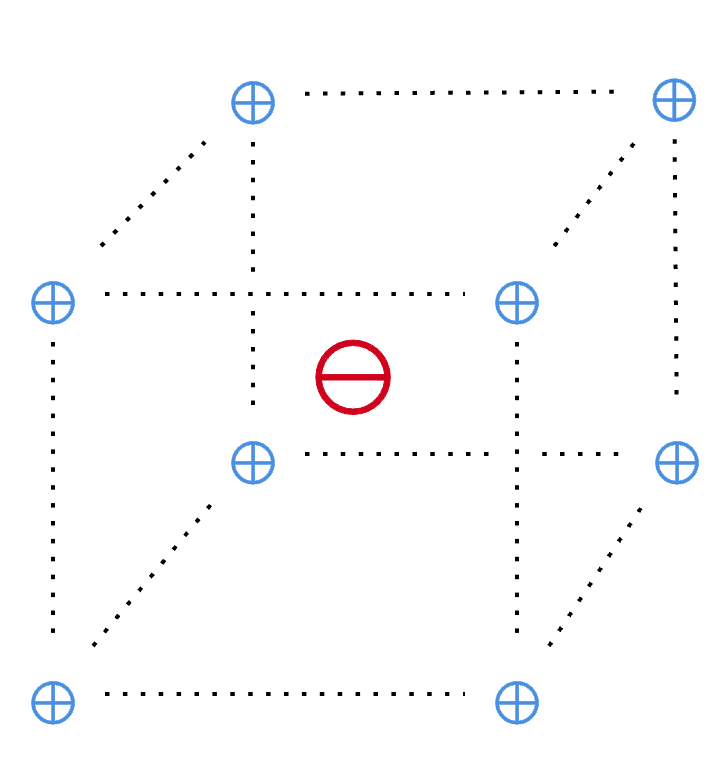}
    \caption{This is the configuration of charges and masses which we refer to as a `crystalline atom'. On the vertices we have positive charges $q$ and on the center we have a negative charge $8 \,q$, so that the whole object is neutral. This is another configuration other than the cube discussed previously, which is in equilibrium.}
    \label{fig:molecule config}
\end{figure}

Solving eqn. (\ref{eqn:NewtonHooke}), We find the equilibrium position of the cube with side length $d$ to be a somewhat complicated formula given by 
\begin{eqnarray}
    d&=&\frac{\sqrt[3]{l^2 \left(\left(18+9 \sqrt{2}+2 \sqrt{3}\right)
   m^2+\left(18-9 \sqrt{2}+2 \sqrt{3}\right)
   q^2\right)}}{3^{2/3} \sqrt[3]{m}}\\
   &=& \# \frac{\sqrt[3]{l^2 m^2}}{\sqrt[3]{m}}+  \# \frac{ \sqrt[3]{l^2 m^2}}{m^{7/3}} q^2 +O(q^3)
\end{eqnarray}
\subsection*{A crystalline atom:}
If the charges are arranged in the manner depicted in Fig. \ref{fig:molecule config}, then upon solving eqn. (\ref{eqn:NewtonHooke}), We find the equilibrium position of the crystalline atom with side length $d$ to be
\begin{eqnarray}
    d&&=\frac{\sqrt[3]{l^2 \left(\left(18+9 \sqrt{2}+10 \sqrt{3}\right) m^2+\left(-18-9 \sqrt{2}+62 \sqrt{3}\right) q^2\right)}}{3^{2/3} \sqrt[3]{m}}\\
    &&=\# \frac{\sqrt[3]{l^2 m^2}}{\sqrt[3]{m}}+  \# \frac{ \sqrt[3]{l^2 m^2}}{m^{7/3}} q^2 +O(q^3)
\end{eqnarray}

\section{Perturbations about empty de Sitter spacetime}
\label{charged formalism}

In this section, we review the Regge-Wheeler formalism used to perturb empty de Sitter spacetime. This formalism is a powerful tool due to the spherical symmetry of the background metric.  Deep in the bulk when $d<<r<<l$, the metric perturbation should reproduce the Newtonian potential, which serves as a boundary condition for perturbations. Such perturbations were used by \cite{Fischler:2024cgm} for static configurations of uncharged objects placed within de Sitter spacetime. This serves as a necessary stepping stone for the later analysis of charged and rotating objects. We then generalize these results for objects whose electric charge is sourced by a stress tensor $T_{\mu \nu}$. 

\subsection*{Regge-Wheeler formalism}
Regge and Wheeler \cite{Regge1957} first studied perturbations of the Schwarzschild spacetime in to analyze its stability. The perturbations about the background spacetime are decomposed into spherical harmonics for two classes of perturbations, axial for odd-parity and polar for even parity. Both sets of perturbations can be formulated with or without time dependence. By exploiting the spherical symmetry of the Schwarzschild background, Regge and Wheeler were able to show that each perturbative mode of Einstein's equations decouples and can be treated individually. One can use their formalism to study perturbations around de Sitter spacetime (with or without a black hole), which has spherical symmetry as was done in \cite{Guven1990,Fischler:2024cgm}.  

We choose to study perturbations about the empty static patch with line element given by
\begin{eqnarray}
    ds^2_{dS}= -\left(1-\frac{r^2}{l^2}\right) dt^2+\left(1-\frac{r^2}{l^2}\right)^{-1} dr^2+r^2 d\Omega^2.
\end{eqnarray}
The full metric is described by the empty de Sitter background plus small perturbations $g'_{\mu\nu}= g_{\mu\nu}^{dS}+h_{\mu\nu}$, with $|h_{\mu \nu}| \ll 1$. Then the linearized Einstein equation governs the dynamics of these perturbations. The first-order Einstein's equations are
\begin{eqnarray}
\label{lin einstein}
    \delta R_{\mu\nu}- \Lambda h_{\mu\nu}=0.
\end{eqnarray}
Due to the spherical symmetry of the background, the perturbations decouple into different modes, which can be treated individually. The time-independent polar perturbations in the Regge-Wheeler gauge are given by
\begin{eqnarray}
h_{\mu\nu}^{(L,M)}=Y_{L,M}(\theta ,\phi )\left(
\begin{array}{cccc}
 H_0^{(L,M)}(r) \left(1-\frac{r^2}{l^2}\right)  & 0 & 0  & 0 \\
 0 & \frac{H_2^{(L,M)}(r) }{1-\frac{r^2}{l^2}} & 0  & 0  \\
 0 & 0 & r^2 K^{(L,M)}(r)  & 0 \\
 0 & 0 & 0 & r^2 \sin ^2(\theta ) K^{(L,M)}(r)  \\
\end{array}.
\right)
\end{eqnarray}
 Using the constraint Einstein equation \eqref{lin einstein}, the polar perturbations are parametrized by just two functions.
\begin{eqnarray}
    H^{(L,M)}(r) \equiv H_2^{(L,M)}(r) \equiv H_0^{(L,M)}(r) \qquad \text{and}\,\,\, K^{(L,M)}(r).
\end{eqnarray}
The dynamical equations and the solutions for these perturbations that arise from Einstein's equations are written in eq (3.7-3.10) of \cite{Fischler:2024cgm}. The constants of integration in the solutions to $H^{(L,M)}(r)$ and $ K^{(L,M)}(r)$ are fixed by matching to the gravitational potential in the Newtonian limit. When we consider charged objects, we must include an EM stress tensor, which acts as a source for the Einstein equation. The Regge-Wheeler formalism is still applicable to non-vacuum solutions. The details of the stress tensor decomposition can be found in Appendix \ref{decomposition_stress}.

The time-independent axial (odd) perturbations in Regge-Wheeler gauge are given by
\begin{eqnarray}
    h_{\mu\nu}^{(L,M)}=\left(
\begin{array}{cccc}
 0  & 0 & -\csc (\theta )\, h_0^{(L,M)} \,Y_{L,M}^{(0,1)}(\theta ,\phi )&\,\,\, \sin (\theta )\, h_0^{(L,M)}\, Y_{L,M}^{(1,0)}(\theta ,\phi )\\
 0 & 0 & -\csc (\theta )\, h_1^{(L,M)} \,Y_{L,M}^{(0,1)}(\theta ,\phi )& \,\,\,\sin (\theta )\, h_1^{(L,M)} \,Y_{L,M}^{(1,0)}(\theta ,\phi )\\
 * & * & 0  & 0 \\
 * & * & 0 & 0  \\
\end{array}
\right)
\end{eqnarray}
Here the $*$ means the symmetric component and $Y_{L,M}^{(1,0)}(\theta ,\phi )\equiv \partial_{\theta} Y_{L,M},\,Y_{L,M}^{(0,1)}(\theta ,\phi )\equiv \partial_{\phi} Y_{L,M}$.
The dynamical equations for axial perturbations and their solutions are discussed later in section \ref{rotating formalism} as they apply to rotating objects. 
\section{Charged objects}
\label{charged objects}
In this section, we study the metric perturbations sourced by charged objects deep within the de Sitter bulk. Such perturbations also can be studied using the Regge-Wheeler formalism discussed above. The novelty in such cases being the presence of an electromagnetic stress tensor. First, we study the perturbations due to a massive object with charge $q$. As expected, this configuration reproduces the Reissner-Nordstr\"{o}m black hole in de Sitter spacetime. Next, we study the perturbations due to an electric dipole with net charge zero with length scale $d$. In the subsequent subsection, we study the deformation of the horizon due to more complicated configurations of charges like those seen figs. \ref{fig:cube config} and \ref{fig:molecule config}.

The linearized Einstein equations sourced by a stress tensor are given by
\begin{equation}
\label{einstein lin}
    \delta R_{\mu \nu}-\frac{3}{l^2}h_{\mu\nu}-\frac{1}{2} \delta R g_{\mu\nu} = 8 \pi G T_{\mu \nu}.
\end{equation}
The solutions to Einsteins equations split into homogeneous and inhomogeneous parts. The homogeneous solutions are the uncharged mass distributions alluded to previously and found in \cite{Fischler:2024cgm}. The homogeneous solutions fix all the constants of integration of the differential equations, while the inhomogeneous solutions capture the charge dependence of the system. We study four configurations of charged objects: a single charged mass, a charged dipole, charges arranged on the vertices of a cube, and a ``crystalline atom'' with the same symmetry group as the cube.

\subsection{A single charged mass}
We consider an object of mass $m$ and charge $q$ placed at the origin of the static patch. We analyze this object using the Regge-Wheeler formalism and show that the perturbed metric indeed recovers the Reissner-Nordstr\"{o}m-de Sitter solution, which describes a charged black hole in de Sitter spacetime. The perturbations of de Sitter spacetime that describe this configuration are given by the spherically symmetric $L=0$ mode. Thus, the perturbations are characterized by only two unknown functions $H_0(r)$ and $K_0(r)$.

To find the electromagnetic stress tensor that sources the perturbations, we first write the electric potential due to a single charge
\begin{eqnarray}
    A_{\mu}=\{A_t,0,0,0\}, \qquad A_t=\frac{Q}{r},
\end{eqnarray}
where we have set $\frac{1}{4 \pi \epsilon_0}=1$. The field strength $F_{\mu\nu}=\partial_{\mu}A_{\nu}- \partial_{\nu}A_{\mu}$ solves Maxwell's equations in de Sitter spacetime $\nabla_{\mu}F^{\mu\nu}=0$ away from the sources. The electromagnetic stress tensor in curved spacetime is given by
\begin{eqnarray}
    T_{\mu\nu}= \Bigg(F_{\mu\alpha}F_{\nu}^{\,\,\alpha}- \frac{1}{4} g_{\mu\nu}F_{\alpha\beta}F^{\alpha\beta}\Bigg),
\end{eqnarray}
which we substitute into the linearized Einstein equations eqn. \eqref{einstein lin}. We label the left hand side of the linearized Einstein equations as $E_{ab}$ and solve  $E_{tt}= 8 \pi G T_{tt}$ and $E_{rr}= 8 \pi G T_{rr}$. These are two independent equations for two unknown functions. 

The $L=0, M=0$ modes of Einstein's equations are written explicitly as 
\begin{eqnarray}
&&r^2 \left(r \left(\left(r^2-l^2\right)
   H'(r)+r (l-r) (l+r) K''(r)+\left(3 l^2-4 r^2\right)
   K'(r)\right)-H(r) \left(l^2-3r^2\right)\right)+\nonumber\\
 && =-l^2 r^2
  K(r)-8 \pi ^{3/2} G l^2 Q^2\nonumber\\
&& -8 \pi ^{3/2} G l^2 Q^2-r^2 \left(r (r-l) (l+r) H'(r)-H(r)
   \left(l^2-3 r^2\right)+r \left(l^2-2 r^2\right)
  K'(r)+l^2 K(r)\right) =0\nonumber.\\
\end{eqnarray}
The constant of integration can be fixed by demanding that these constants reproduce the correct Newtonian potential for the object. After fixing the constants, the metric perturbation due to a single charged object is
\begin{eqnarray}
    h_{tt}=\frac{2 G m r}{l^2}+\frac{2 G m}{r}-\frac{G Q^2}{r^2},\quad h_{\theta\theta}=2  G\, m\, r
\end{eqnarray}
We can compare our results to a Reissner-Nordstr\"{o}m-de Sitter black hole by redefining the $r$ coordinate as $r \rightarrow r \sqrt{\frac{2m}{r}+1}$, which we transform to match the angular portions for both metrics. 
Then the metric components $g_{tt}=g_{tt}^{dS}+h_{tt}$ and $g_{\theta\theta}=g_{\theta\theta}^{dS}+h_{\theta\theta}$ become
\begin{eqnarray}
    g_{tt}= \frac{2 G m}{r}-\frac{G
   Q^2}{r^2}+\frac{r^2}{l^2}-1, \quad g_{\theta\theta}=r^2
\end{eqnarray}
The horizon can be found by finding where $g_{tt}=0$. We find the the outermost horizon to be located at \footnote{Here we have redefined $m \rightarrow G m$ and $ G Q^2 \rightarrow Q^2$} 
\begin{eqnarray}
    r_h=l-  m +\frac{  Q^2}{2 l}, 
\end{eqnarray}
which is the known cosmological horizon of the Reissner-Nordstr\"{o}m-de Sitter spacetime. The perturbed metric we find indeed reproduces the Reissner-Nordstr\"{o}m black hole in de Sitter spacetime, which is described by the line element
\begin{eqnarray}
    ds^2=-f(r) dt^2+\frac{dr^2}{f(r)}+r^2 (d\theta^2+\mathrm{sin}^2 \theta d \phi^2), \quad f(r)= 1- \frac{2 G m}{r}+\frac{ Q^2}{r^2}-\frac{r^2}{l^2}.
\end{eqnarray}

\subsection{A dipole}
Next, we consider a dipole which consists of positive and negative charges with mass $m$ separated by a distance $d$ along the $z$-axis. The dipole moment for the system is $\vec{p}=q d\,  \hat{k}$. The electric potential due to a dipole in the de Sitter spacetime is \footnote{The potential in dS spacetime has an additional factor of $(1-r^2/l^2)$ compared to flat spacetime. }
\begin{eqnarray}
    A_t= \frac{p\, \mathrm{cos}\, \theta}{r^2}(1-\frac{r^2}{l^2}),\qquad  A_i=0
\end{eqnarray}
The field strength can be calculated as
\begin{eqnarray}
F^{tr}=\frac{-2\, p  \,\mathrm{cos} \theta}{r^3}, \quad F^{t \theta}= \frac{-p \,\mathrm{ sin}\, \theta}{r^4}, \quad F^{t \phi}=0.
\end{eqnarray}
The field strength is a solution of the Maxwell equation (in curved space) away from the sources, and its EM stress tensor is covariantly conserved $\nabla_{\mu}T^{\mu\nu}=0$. The exact expression for the stress tensor is complicated and uninformative to write down. We decompose the stress tensor into tensor harmonics following the decomposition in the Appendix \ref{stress_decom}. Apart from $L=0$, the first non-zero mode that appears is the $L=2$ mode. We solve the Einstein equation for each $L$ separately. As we have placed the dipole along the $z$ axis, only $M=0$ will contribute (a consequence of the azimuthal symmetry).\\

 \textbf{Area of the horizon:-} We now calculate the area of the perturbed cosmological horizon. The determinant of the induced metric on the horizon in the linear order in perturbation is given by
 \begin{eqnarray}
    \sqrt{|g|}= r^2 \,\mathrm{sin}\, \theta \Big(1+\sum_{L,M} K^{(L,M)}(r) Y_{LM}(\theta,\phi) \Big)
 \end{eqnarray}
 Only the $L=M=0$ mode contributes to the total area of the horizon. The higher $L$ modes deform the horizon's shape but in an area-preserving manner.\\

\textbf{L=0 modes:-} We start with $L=M=0$ modes of the Einstein equation. The constants of the in-homogeneous solutions are fixed by demanding the $L=0$ mode should reproduce the gravitational potential due to dipole (see eq 5.2 and 5.3 of \cite{Fischler:2024cgm} for fixing these constants). After fixing these constants, the complete solution to the inhomogeneous equation can be written as
\begin{eqnarray}
    h(r)=\frac{8 \pi ^{3/2} d^2 q^2 \left(3 l^2-2 r^2\right)}{9 r^4
   \left(r^2-l^2\right)}+\frac{4 \sqrt{\pi } m
   \left(l^2+r^2\right)}{r \left(l^2-r^2\right)}, \quad K(r)=\frac{4 \sqrt{\pi } m}{r}-\frac{8 \pi ^{3/2} d^2 q^2}{9 r^4}
\end{eqnarray}
These solutions change the location and area of the would-be horizon. Higher modes will affect only the shape of the horizon. The location and shape of the horizon will be discussed at the end of this section.\\

 \textbf{L=1:} There is no projection of the stress tensor in $L=1$ mode. The homogeneous solution also vanishes.\\
 
\textbf{L=2:}  The inhomogeneous equation can be solved in the multipole region ($d<<r<<l$). The solutions, after comparing them with Newtonian potential, fix the constant of integration\footnote{It is interesting to see that the constants are fixed only by the mass-dependent term in the Newtonian potential.}. 
\begin{eqnarray}
 && H^{(2,0)}(r)Y_{2,0}(\theta,\phi)=\frac{d^2 (3 \cos (2 \theta )+1) \left(9 m r-16 \pi  q^2\right)}{18 r^4} ,\\
 &&  K^{(2,0)}(r)Y_{2,0}(\theta,\phi)=\frac{d^2 (3 \cos (2 \theta )+1) \left(3 m r-2 \pi  q^2\right)}{6 r^4}\nonumber
\end{eqnarray}
Having fixed the constant, we now solve the Einstein equation near the horizon $r=l$. The particular solution near $r=l$ is
\begin{eqnarray}
    H^{(2,0)}(r)=\frac{16 \pi ^{3/2} d^2 q^2 \left(r^2-l^2\right)}{9 \sqrt{5} l^6},\quad  K^{(2,0)}(r)=\frac{32 \pi ^{3/2} d^2 q^2}{3 \sqrt{5} l^4}
\end{eqnarray}
There are no higher $L>2$ modes of stress tensor for this case. This completes the discussion of solutions to Einstein's equation. \\

To compare with the Reissner-Nordstr\"{o}m-de Sitter solution, we redefine the radial coordinate to be
 \begin{eqnarray}
    r'= r \sqrt{\frac{2 m}{r}+1} \sim r-m
 \end{eqnarray}
 With these $L=0$ and $L=2$ perturbations, we find the location of the horizon by solving $g_{tt}=g_{tt}^{dS}+h_{tt}=0$. \\
 
\begin{figure}
     \centering
     \includegraphics[width=0.5\linewidth]{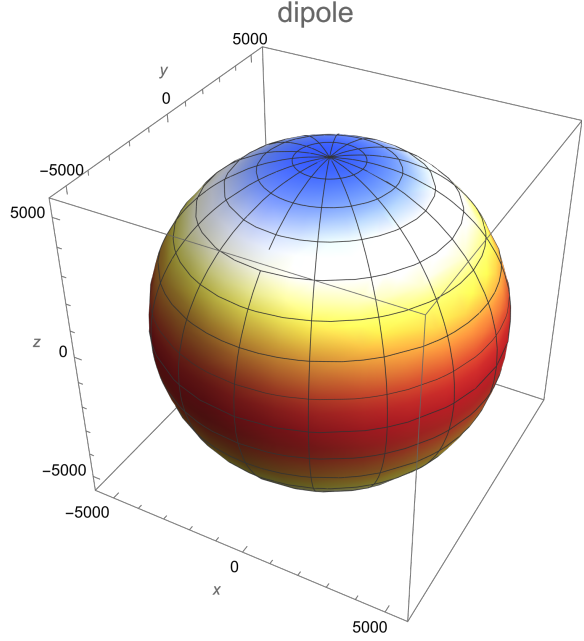}
     \caption{We have plotted the shape of the cosmological horizon when we have a dipole in the bulk. The shape is described by eq \eqref{horizon dipole}. We took the parameters $l=10000,\, m=2000,\, q=4000,\, d=2000$ away from their regime of validity to portray the features (dips and protrusions) on the horizon. At the North Pole and South Pole, the horizon dips (shown in blue). Near the equator, the horizon protrudes (shown in red). The dip and protrusions are defined relative to the $L=0$ mode of the perturbation. In summary, the cosmological horizon dips and protrudes in an area-preserving manner.}
     \label{fig:dipole plot}
 \end{figure}

  \textbf{Location and shape of the horizon }
The location of the horizon can be found by using the following ansatz for the horizon location,
 \begin{eqnarray}
     r'_H=l-m- \epsilon(\theta,\phi),
 \end{eqnarray}
 which we substitute into the horizon location condition $g_{tt}=g_{tt}^{dS}+h_{tt}^{(0,0)}+h_{tt}^{(2,0)}=0$. We solve this equation to linear order in $\epsilon (\theta,\phi)$. This gives the shape of the horizon.
 \begin{eqnarray}
 \label{horizon dipole}
     r_H'&&=l-m-(\frac{2
   m^2}{l}+\frac{8 m^3}{l^2}+\frac{32
   m^4-\frac{2}{9} \pi  d^2 q^2}{l^3}+\frac{4 m \left(3 d^2 m^2 \cos (2 \theta
   )+d^2 \left(m^2-\pi  q^2\right)+34 m^4\right)}{l^4})\nonumber\\
   &&+\frac{16 m^2 \left(3 d^2 \cos (2 \theta ) \left(18 m^2-\pi 
   q^2\right)+2 d^2 \left(9 m^2-10 \pi  q^2\right)+333
   m^4\right)}{9 l^5} +O(\frac{1}{l^6}).
 \end{eqnarray}
In the first line, we have included the higher correction coming from the $L=0$ mode (3rd, 4th, and 5th term). These terms don't have any angular dependence. The angular dependence of the charges is present at $1/l^4$ order and higher.  In fig. \ref{fig:dipole plot}, we have plotted the shape of the horizon. The horizon dips and protrudes in a total-area preserving manner. To understand the detailed aspects of the horizon, we calculate the area of one of the dips near the north pole (shown in blue in fig. \ref{fig:dipole plot}). The dip and protrusions on the horizon are defined relative to the area of the $L=0$ mode. We instead choose an angle $\theta_{crit}$ which roughly corresponds to where the horizon reaches the value given by only its $L=0$ perturbation. We find the dip near the north pole is spanned by the region $\phi \in [0,2 \pi], \, \theta \in [0, \theta_{crit}]$, where $ \theta_{crit}\sim \pi/4$. \\
We integrate the area element of the dip region and subtract the area of the $L=0$ mode, giving the difference in area 
\begin{eqnarray}
\Delta A\sim-\frac{8 \sqrt{2} \pi  d^2 m^3 (l-m)}{l^4}\leq 0
\label{dip}
\end{eqnarray}
where we have suppressed the term at the order of $O(1/l^4)$ and higher. We find that the dip of the cosmological horizon near the north pole has less area than the $L=0$ mode assuming $q << m$. Similarly, one can find the difference in the area near the equator and find that it is greater than the area of the cosmological horizon only due to $L=0$ mode \footnote{The equilibrium position $d$ is fixed in eq \eqref{dipole equi}}. 
\begin{eqnarray}
\label{portrusion}
\Delta A \sim \frac{16 \sqrt{2} \pi  d^2 m^3 (l-m)}{l^4} \geq 0.
\end{eqnarray}
The areas of the dip \eqref{dip} (there are two dip regions, one near the North pole and one near the south pole) and protrusion regions \eqref{portrusion} cancel, so the total area of the cosmological horizon is preserved under every $L>0$ perturbations. \\
\textbf{The electric flux on the horizon:}\\
The electric field at the horizon can be written as
  \begin{eqnarray}
     F^{tr}=E^r=-\frac{2 p \cos (\theta )}{l^3}+O(\frac{1}{l^4}), \quad F^{t\theta}=E^{\theta}=-\frac{p \sin (\theta )}{l^4}+O(\frac{1}{l^5}).
 \end{eqnarray}
 
\begin{figure}
    \centering
    \includegraphics[width=0.65\linewidth]{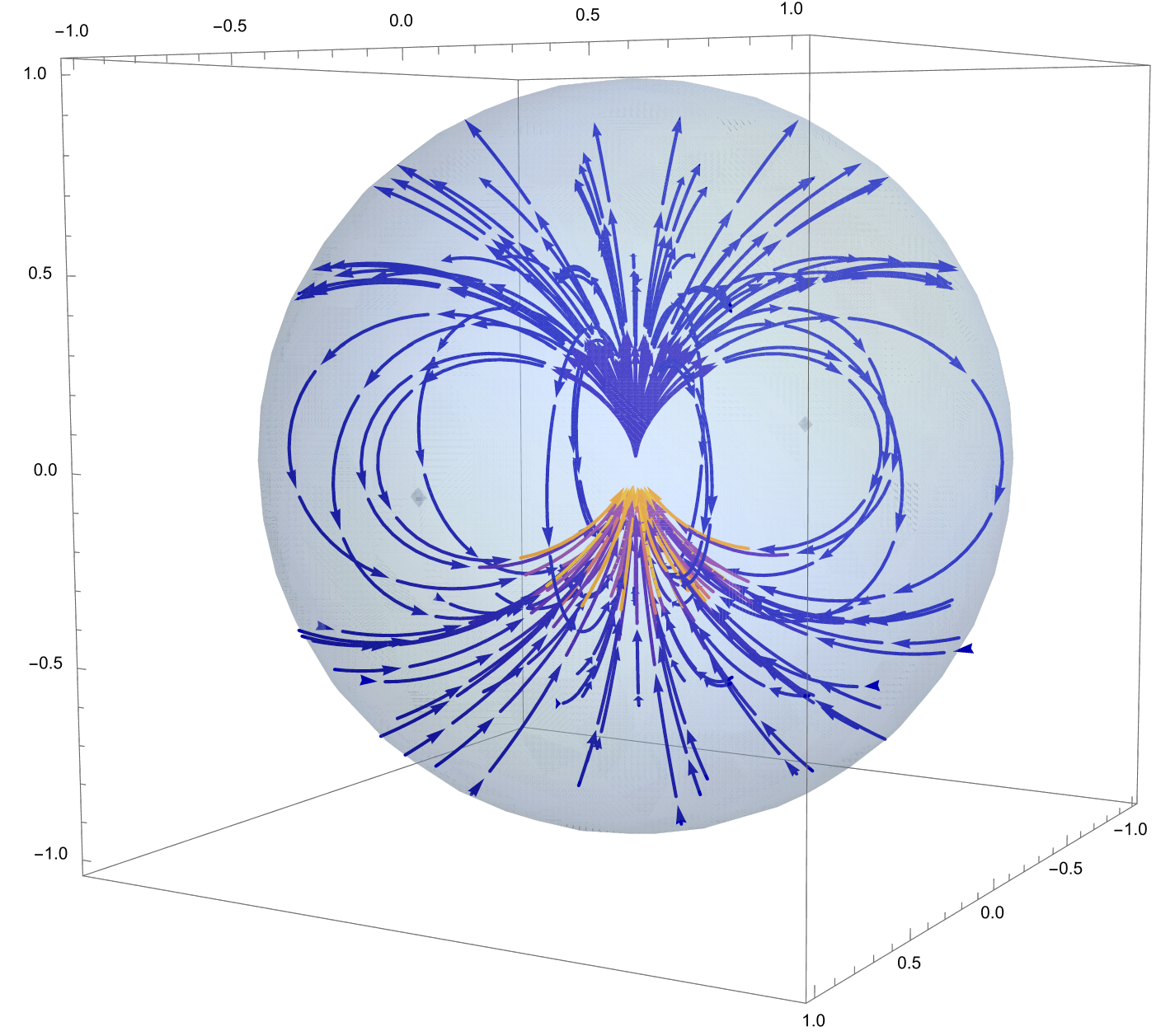}
    \caption{We have plotted the field lines of the dipole placed in the bulk. The field lines end on the cosmological horizon shown here as a shaded surface. The field lines are consistent with the charges induced on the cosmological horizon by Gauss' law.} 

    \label{fig:field lines}
\end{figure}

Electric field lines that originate from the dipole in the bulk end at the cosmological horizon. We schematically show these field lines in fig. \ref{fig:field lines}. These field lines are indeed consistent with the induced smeared charge at the horizon.

We use the Komar formula \cite{Misner:1973prb,Carroll:2004st} to find electric charges on the horizon
 \begin{eqnarray}
 \label{komar}
     Q= -\int_{\partial \Sigma} d^2 x \sqrt{\gamma^{(2)}} n_{\mu}\sigma_{\nu} F^{\mu\nu},
 \end{eqnarray}
where the Cauchy slice $\Sigma$ is the above formula is constant $t$ slice of the empty de Sitter spacetime. Hence, the normal vector $n_{\mu}=\{n_0,0,0,0\}$ is along the time direction. The $\partial \Sigma$ is $S^2$ of the empty de Sitter. Then the normal vector to the sphere is denoted by $\sigma_{\mu}$ and it is the radial direction\footnote{The original cosmological horizon with $r_h=l$ is not null when we add the masses in the bulk. This surface is like an ordinary space-like surface where the Komar formula can be applied. The normal vector $n^{\mu}$ is the unit normal to the $\Sigma$. Other outward-pointing normal vector is $\sigma^{\mu}$.}. The imprint of the perturbations can be found in the induced metric $\gamma^{(2)}_{ab}$ and field strength $F^{tr}$. Explicitly, we have
 \begin{eqnarray}
     &&n_0=-(1-\frac{r^2}{l^2})^{\frac{1}{2}},\quad  n_i=0, \quad n_{\mu}n^{\mu}=-1\nonumber\\
     &&\sigma_1= (1-\frac{r^2}{l^2})^{\frac{-1}{2}}, \quad \sigma_0=\sigma_2=\sigma_3=0,\quad \sigma_{\mu}\sigma^{\mu}=1
 \end{eqnarray}
Using the Komar formula to find the induced charge through the dip (near the north pole) 
 \begin{eqnarray}
 \label{charge north pole}
     Q=\int_{0}^{\theta_{crit}}\int_0^{2 \pi} d \phi \,d \theta \, \, r_h^{2}\, \mathrm{sin}\, \theta F^{tr}=&&- \frac{2 \pi q d}{l} \int_{0}^{\theta_{crit}} \mathrm{sin} 2\theta +O(\frac{1}{l^2})\nonumber\\
     =&&-\frac{ \pi  d q }{l}+O(\frac{1}{l^2})
 \end{eqnarray}
 In the dip region near the south pole, one finds a positive smeared charge. We see that the horizon has a net zero charge but becomes polarized due to the charges deep within the bulk.
 
The dipole solution can be compared with the Reissner-Nordstr\"{o}m-de Sitter solution, with electric field $F^{tr}=-\frac{q}{r^2}$. On the cosmological horizon, the charge of Reissner-Nordstr\"{o}m-de Sitter is given by
 \begin{eqnarray}
     Q_{RN}=\int_{0}^{\theta_{crit}}\int_0^{2 \pi} d \phi \,d \theta \, \, r_h^{2}\, \mathrm{sin}\, \theta F^{tr}=- 2 \pi q \int_{0}^{\theta_{crit}} \mathrm{sin} \theta= -0.29 \pi q.
 \end{eqnarray}
Thus, from horizon data, we can measure quantitative differences between the two geometries. 

In summary, the induced electric charge on the horizon contains information of the objects within the bulk. From the area of a dip/protrusion, the equilibrium distance, $d$, of the objects in the bulk can be read off. The sign of induced charges on the horizon also determines the corresponding distribution of charges within the bulk. Hence, the area and induced charges have all the pieces of information about the object that was placed in the bulk. 
 
\subsection{A cube}
Next, we study a cube where the placement of positive and negative charges is shown in fig. \ref{fig:cube config}. In this configuration, we alternatively place positive and negative charges along the face so that there is no net charge on any face. This configuration of charge has the same symmetry group as the mass configuration. The Coulomb potential for this configuration can be written as
 \begin{eqnarray}
     A_t=\frac{60 d^3 q \sin ^2(\theta ) \cos (\theta ) \sin (2 \phi )}{r^4}(1-\frac{6 r^2}{5 l^2}+\frac{r^4}{5 l^4}), \quad A_i=0
 \end{eqnarray}
 The field strength for the above potential is
 \begin{eqnarray}
    && F^{tr}=\frac{48  d^3 q \sin ^2(\theta ) \cos (\theta ) \left(3 r^2-5 l^2\right) \sin (2 \phi )}{l^2 r^5}\nonumber\\
    && F^{t\theta}=\frac{6  d^3 q \sin (\theta ) (3 \cos (2 \theta )+1) \left(5 l^2-r^2\right) \sin (2 \phi )}{l^2 r^6}\nonumber\\
    && F^{t \phi}=-\frac{24  d^3 q \cos (\theta ) \left(r^2-5 l^2\right) \cos (2 \phi )}{l^2 r^6}
 \end{eqnarray}
 \textbf{Solving Regge-Wheeler equation:}\\
 
 As in the case of the dipole, we solve the Regge-Wheeler equation with a stress tensor sourced by electric charges.\\
 
 \textbf{L=0:}\\
We find the Einstein and stress tensor in the $L=0$ sector following the discussion in appendix \eqref{stress_decom}. We solve $G_{tt}= 8 \pi G T_{tt}$ and $G_{rr}= 8 \pi G T_{rr}$ projected onto the $L=M=0$ sector. The constants of integration can be found by matching with the Newtonian potential. The complete solution is 
 \begin{eqnarray}
    && H^{(0,0)}(r)= -\frac{1536 \pi ^{3/2} d^6 q^2 \left(175 l^6-380 l^4 r^2+266 l^2 r^4-28 r^6\right)}{245 l^4 r^8 \left(l^2-r^2\right)}-\frac{4 \sqrt{\pi } m
   r}{r^2-l^2}-\frac{4 \sqrt{\pi } l^2 m}{r \left(r^2-l^2\right)}\nonumber\\
  && K^{(0,0)}(r)=-\frac{1536 \pi ^{3/2} d^6 q^2 \left(75 l^4-56 l^2 r^2+14 r^4\right)}{245 l^4 r^8}+\frac{4 \sqrt{\pi } m}{r}
 \end{eqnarray}
There are no $L=1,2,3$ modes for this configuration.\\
 \textbf{L=4}\\
 Next, we solve the Einstein equation in $L=4,M=0,\pm 4$ sector. The inhomogeneous equation can be solved in the multipole region ($d<<r<<l$). The solutions, after comparing them with Newtonian potential, fix the constant of integration\footnote{It is interesting to see that the constants are fixed only by the mass-dependent term in the Newtonian potential.}. 
 \begin{eqnarray}
    && H^{(4,0)}(r)=\frac{181760 \pi ^{3/2} d^6 q^2}{187 r^8}-\frac{14 \sqrt{\pi } d^4
   m}{3 r^5}, \quad K^{(4,0)}(r)=\frac{74240 \pi ^{3/2} d^6 q^2}{187 r^8}-\frac{14 \sqrt{\pi } d^4
   m}{3 r^5}\nonumber\\
    && H^{(4,\pm4)}(r)=\frac{90880 \sqrt{\frac{10}{7}} \pi ^{3/2} d^6 q^2}{187
   r^8}-\frac{\sqrt{70 \pi } d^4 m}{3 r^5},\quad K^{(4,\pm 4)}(r)= \frac{37120 \sqrt{\frac{10}{7}} \pi ^{3/2} d^6 q^2}{187
   r^8}-\frac{\sqrt{70 \pi } d^4 m}{3 r^5}\nonumber\\
 \end{eqnarray}
 Next, we solve the equation near the horizon $r=l$ and the particular solution can be written as 
 \begin{eqnarray}
   &&  H^{(4,0)}(r)=-\frac{24576 \pi ^{3/2} d^6 q^2 \left(r^2-l^2\right)}{275 l^{10}}, \quad K^{(4,0)}(r)=\frac{32768 \pi ^{3/2} d^6 q^2}{165 l^8}\nonumber\\
       && H^{(4,\pm 4)}(r)= \frac{12288 \sqrt{\frac{2}{35}} \pi ^{3/2} d^6 q^2 \left(r^2-l^2\right)}{55 l^{10}}, \quad K^{(4,\pm 4)}(r)=-\frac{16384 \sqrt{\frac{2}{35}} \pi ^{3/2} d^6 q^2}{33 l^8}
 \end{eqnarray}

\begin{figure}
     \centering
     \includegraphics[width=0.5\linewidth]{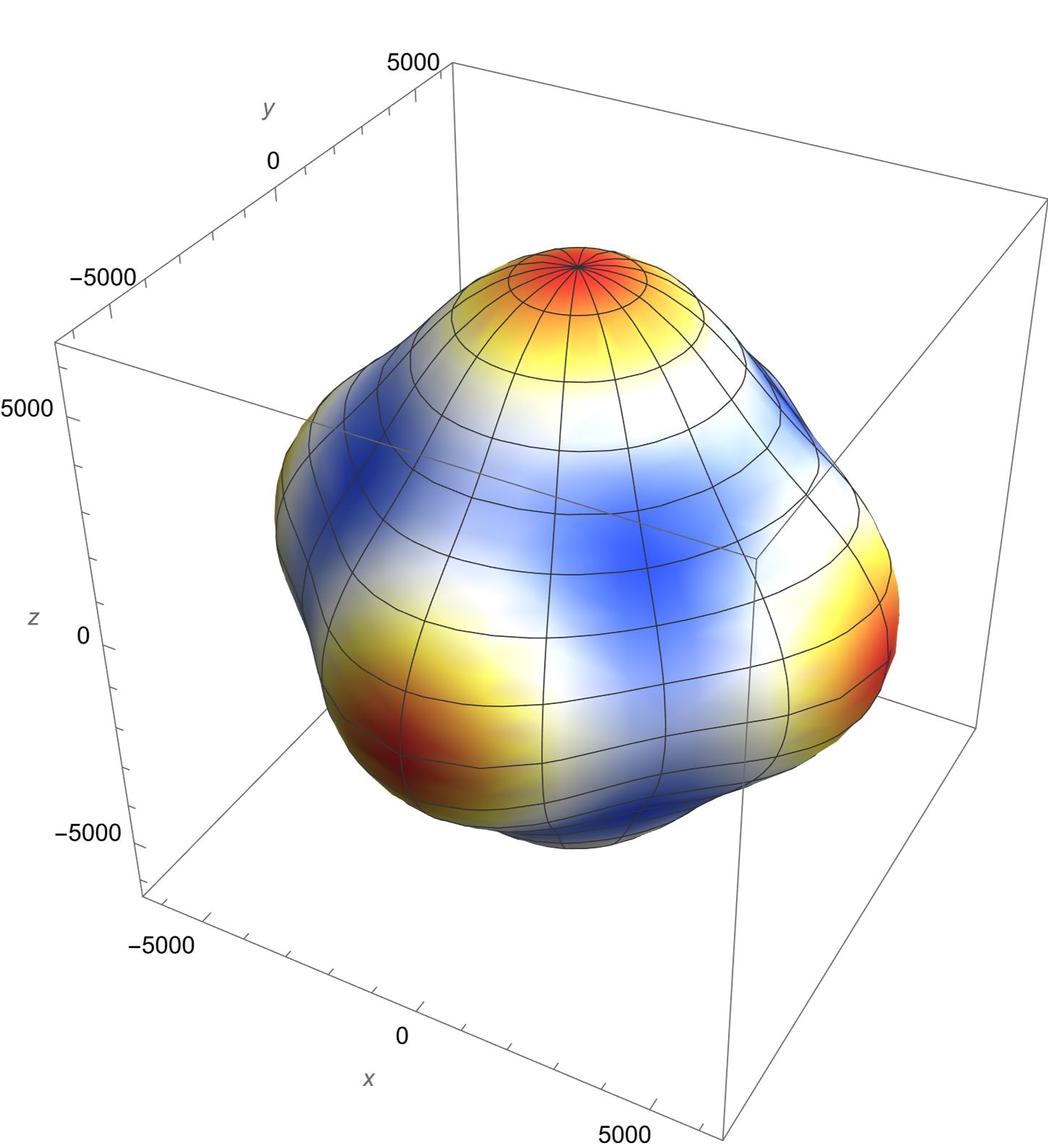}
     \caption{The cosmological horizon for a cube placed within the bulk. The horizon's shape is described by eq \eqref{horizon cube},  with parameters $l=10000,\, m=2000,\, q=2000,\, d=5000$, which are far outside the range of perturbative validity but highlight geometric features. The coloring of the horizon is represented by a temperature map where blue corresponds to dips of the horizon and red corresponds to protrusions. The dip and protrusions are defined relative to the $L=0$ mode of the perturbation. The shape of the horizon is the dual Platonic solid to a cube, the octahedron. We see here that the charges do not change the geometry of the cosmological horizon but rather enhance its features.}
     \label{fig:cube diff}
 \end{figure}

To find the horizon, we solve $g_{tt}=g_{tt}^{dS}+h_{tt}^{(0,0)}+h_{tt}^{(4,0)}+h_{tt}^{(4,4)}+h_{tt}^{(4,-4)}=0$. As before, only $L=0$ mode can change the area of the horizon. The higher modes can only change the shape of the horizon in a manner that preserves the area. We take the ansatz $r_h'=l-m- \epsilon(\theta,\phi)$ for the horizon. For this configuration of charges and masses, the location and shape of the horizon (and plotted in fig. \ref{fig:cube diff}) is
 \begin{eqnarray}
 \label{horizon cube}
     r_h'=&&l-m-\Bigg(\frac{2 m^2}{l}+\frac{8
   m^3}{l^2}+\frac{32 m^4}{l^3}+\frac{136 m^5}{l^4}+\frac{592 m^6}{l^5}\nonumber\\
   &&-\frac{3 m^3 \left(5 d^4 \left(8 \sin ^4(\theta ) \cos (4 \phi )+4
   \cos (2 \theta )+7 \cos (4 \theta )\right)+9 \left(d^4-384
   m^4\right)\right)}{4 l^6}\nonumber\\
&&+\frac{\frac{-12672}{245} \pi  d^6 q^2-\frac{95}{2} d^4 m^4 \left(8 \sin ^4(\theta ) \cos (4 \phi )+4 \cos (2 \theta )+7 \cos (4 \theta
   )\right)-\frac{171 d^4 m^4}{2}+11360 m^8}{l^7}\Bigg)\nonumber\\
 \end{eqnarray}
In the first line, we have written the contribution of the higher-order terms of monopole perturbation ($L=0$). These don't have any angular dependence as expected from the monopole contribution. In the second line, we have contributions from the $L=4$ modes which do have the angular dependence. The contribution of the charges only appears at $O(1/l^7)$ order. The Electric field $E^r$ at the horizon is
 \begin{eqnarray}
     F^{tr}=-\frac{96  d^3 q \sin ^2(\theta ) \cos (\theta ) \sin (2 \phi )}{l^5}+ O(\frac{1}{l^6})
    \label{eqn: cube fields}
 \end{eqnarray}
 As in the dipole case, the field lines are consistent with the induced charge on the horizon. \\
 
 \textbf{The area and flux from the cap region:} The dip and protrusions on the horizon are defined relative to the area of the $L=0$ mode. We find the cap near the north pole is spanned by the region $\phi \in [0,2 \pi], \, \theta \in [0, \theta_{crit}]$, where $ \theta_{crit}\sim \pi/6$. Now we calculate the difference in the area of the top cap region relative to the area due to the monopole term. 
 \begin{eqnarray}
\Delta A\sim \frac{27 \sqrt{3} \pi  d^4 m^3 (l-m)}{4 l^6} \geq 0
 \end{eqnarray}
 Here we have suppressed the term of the $O(1/l^7)$ and higher, but it won't make any qualitative difference.
 The charges are more interesting and they can be calculated using the Komar formula \eqref{komar} as before
 \begin{eqnarray}
     Q[\phi]=\int_{0}^{\theta_{crit}} \,d \theta \, \, r_h^{2}\, \mathrm{sin}\, \theta F^{tr}\sim -\frac{3  d^3 q (l-m)^2 \sin (2 \phi )}{2 l^5}
 \end{eqnarray}
 Now we can integrate over different values of $\phi$ to see the effect of individual charges.
 \begin{eqnarray}
    Q_{q}=\int_{0}^{\pi/4} Q[\phi]= -\frac{3  d^3 q (l-m)^2}{4 l^5}, \quad Q_{2 q}= \int_{0}^{\pi/2} Q[\phi]= -\frac{3  d^3 q (l-m)^2}{2 l^5}
\end{eqnarray}
The charge $Q_q$ is due to the flux of one of the positive charges $q$. The charge $Q_{2q}$ is the contribution due to the flux of the two charges $2 q$. The contribution of the next negative charges can be understood when we integrate over $\phi \in [\pi/2,3 \pi/4]$
\begin{eqnarray}
    Q_{-q}=\int_{\pi/2}^{3\pi/4} Q[\phi]= \frac{3  d^3 q (l-m)^2}{4 l^5}, \quad  Q_{-2q}=\int_{\pi/2}^{\pi} Q[\phi]= \frac{3  d^3 q (l-m)^2}{2 l^5},
\end{eqnarray}
 The total charge through the entire top region is zero. Drawing the field lines in this case is messier, but the induced charges are consistent with the field lines.
 \subsection{A crystalline atom}
\begin{figure}
     \centering
     \includegraphics[width=0.50\linewidth]{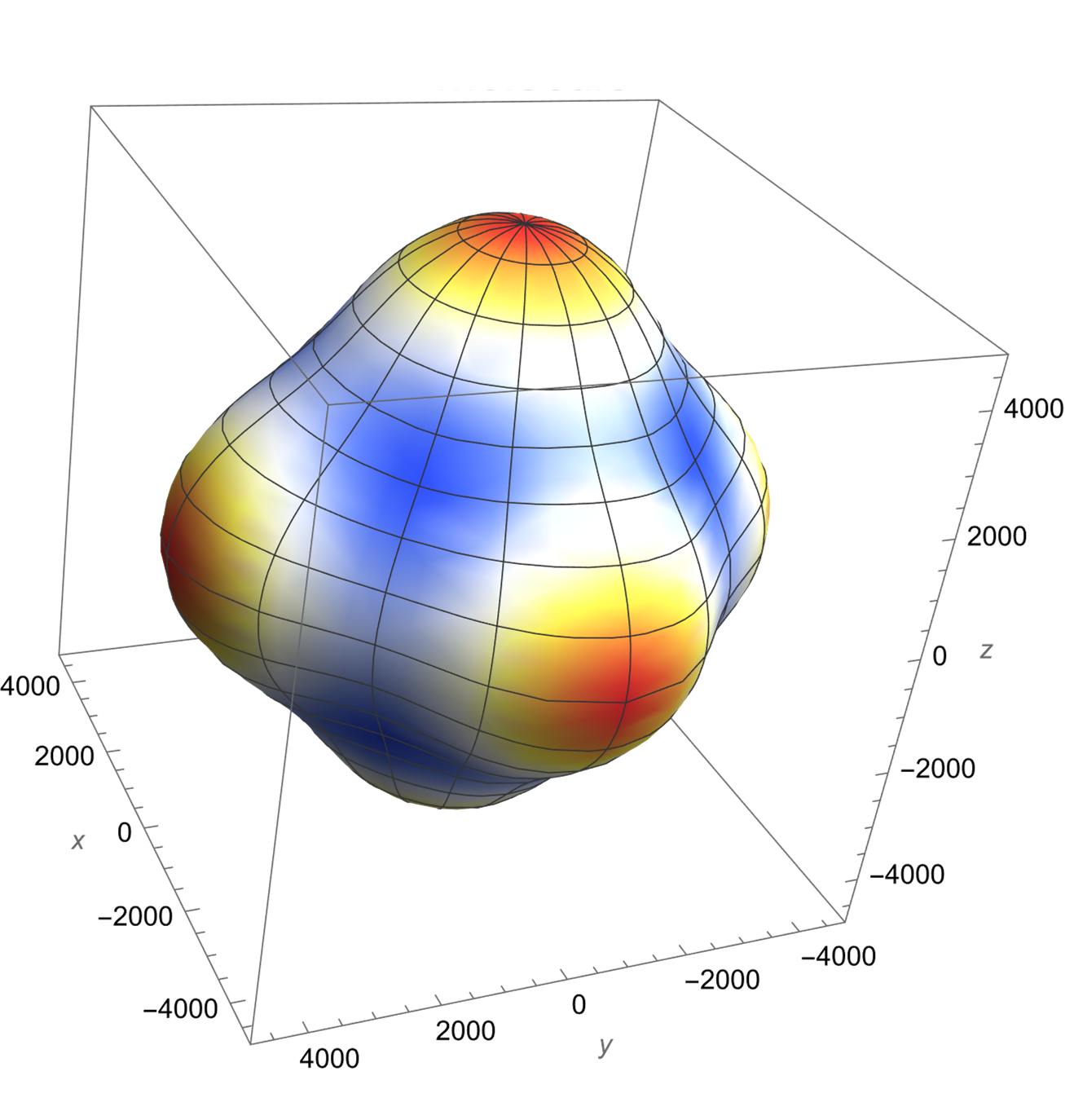}
    \caption{We have plotted the shape of the cosmological horizon when we have a `crystalline atom' in bulk. The shape is described by eq \eqref{molecule shape}. We took the parameters $l=10000,\, m=1000,\, q=1000,\, d=4000$ away from their regime of validity to portray the dips and protrusions on the horizon. The horizon has the shape of the dual object of the cube-octahedron. The coloring of the horizon is represented by a temperature map where blue corresponds to dips of the horizon and red corresponds to protrusions. The shape of the horizon is the dual Platonic solid to a cube, the octahedron. Near the north pole, the yellow region we call it `cap'. The dip and protrusions are defined relative to the $L=0$ mode of the perturbation.}
     \label{fig:molecule plot}
 \end{figure}

 In the next case, the arrangement of constituents (with mass and charges) is shown in Fig. \ref{fig:molecule config}. In the center, we have $-8 q$, and on the vertices, we have $+q$ charges all having the same mass (the total mass of the whole system is $ 2m $). When we set $q=0$, then the configuration is equivalent to the superposition of a single mass and cube solution of \cite{Fischler:2024cgm}. The electric potential with charges can also be superposed in the same manner. The potential is
 \begin{eqnarray}
     A_t=\left(\frac{3 r^4}{7 l^4}-\frac{10 r^2}{7 l^2}+1\right)
   \left(-\frac{35 d^4 q e^{-4 i \phi } \sin ^4(\theta
   )}{4 r^5}-\frac{35 d^4 q e^{4 i \phi } \sin ^4(\theta
   )}{4 r^5}-\frac{7 d^4 q (20 \cos (2 \theta )+35 \cos (4
   \theta )+9)}{16 r^5}\right)\nonumber\\
 \end{eqnarray}
 The monopole term of the potential got canceled, and we are left with $L=4, M=0, \pm 4$ modes\footnote{Not to be confused with the mode decomposition of the stress tensor.}. The field strength can be written as
 \begin{eqnarray}
     &&F^{tr}=\frac{d^4 q e^{-4 i \phi } \left(35 l^4-30 l^2 r^2+3
   r^4\right) \left(20 e^{8 i \phi } \sin ^4(\theta )+20
   e^{4 i \phi } \cos (2 \theta )+35 e^{4 i \phi } \cos (4
   \theta )+20 \sin ^4(\theta )+9 e^{4 i \phi }\right)}{16
   l^4 r^6}\nonumber\\
   &&F^{t \theta}=-\frac{5 d^4 q e^{-4 i \phi } \left(7 l^2-3 r^2\right)
   \left(4 \left(1+e^{8 i \phi }\right) \sin ^3(\theta )
   \cos (\theta )-e^{4 i \phi } (2 \sin (2 \theta )+7 \sin
   (4 \theta ))\right)}{4 l^2 r^7}\nonumber\\
   &&F^{t \phi}=-\frac{5 i d^4 q e^{-4 i \phi } \left(-1+e^{8 i \phi
   }\right) \sin ^2(\theta ) \left(7 l^2-3 r^2\right)}{l^2
   r^7}\nonumber\\
 \end{eqnarray}
 Again, we solve the Regge-Wheeler equation with a stress tensor caused by these electric charge configurations (see fig. \ref{fig:molecule config}).\\
  \textbf{L=0:}\\
 We solve Einstein's equation $G_{tt}= 8 \pi G T_{tt}$ and $G_{rr}= 8 \pi G T_{rr}$ when projected to $L=M=0$ sector. The constants of integration can be found by matching with the Newtonian potential (but now with an object having mass $2m$). The complete solution is 
\begin{eqnarray}
    &&H^{(0,0)}(r)=-\frac{512 \pi ^{3/2} d^8 q^2 \left(441 l^8-1190 l^6
   r^2+1134 l^4 r^4-432 l^2 r^6+81 r^8\right)}{189 l^6
   r^{10} \left(l^2-r^2\right)}-\frac{8 \sqrt{\pi } m
   r}{r^2-l^2}-\frac{8 \sqrt{\pi } l^2 m}{r
   \left(r^2-l^2\right)}\nonumber\\
  && K^{(0,0)}(r)=\frac{8 \sqrt{\pi } m}{r}-\frac{1024 \pi ^{3/2} d^8 q^2
   \left(98 l^4-135 l^2 r^2+54 r^4\right)}{189 l^4 r^{10}}
\end{eqnarray} 
There are no $L=1,2,3$ modes for this configuration.\\
  \textbf{L=4:}\\
 Next, we solve the Einstein equation in $L=4,M=0,\pm4$ sector. In the multipole region, the solutions upon matching with the Newtonian potential are
 \begin{eqnarray}
 &&  H^{(4,0)}=-\frac{1860096 \pi ^{3/2} d^8 q^2}{2717 r^{10}} -\frac{14 \sqrt{\pi } d^4
   m}{3 r^5}, \quad K^{(4,0)}(r)=-\frac{731136 \pi ^{3/2} d^8 q^2}{2717 r^{10}}-\frac{14 \sqrt{\pi } d^4
   m}{3 r^5}\nonumber\\  
&&H^{(4,\pm4)}(r)=-\frac{132864 \sqrt{70} \pi ^{3/2} d^8 q^2}{2717 r^{10}}-\frac{\sqrt{70 \pi } d^4 m}{3 r^5},\quad K^{(4,\pm 4)}(r)= -\frac{52224 \sqrt{70} \pi ^{3/2} d^8 q^2}{2717 r^{10}}-\frac{\sqrt{70 \pi } d^4 m}{3 r^5}\nonumber\\
 \end{eqnarray}
 Upon fixing the integration constant, we now solve the Einstein equation near $r=l$. The particular solution near $r=l$ is
\begin{eqnarray}
  &&  H^{(4,0)}=-\frac{24576 \pi ^{3/2} d^8 q^2 \left(r^2-l^2\right)}{143
   l^{12}},\quad K^{(4,0)}=\frac{32768 \pi ^{3/2} d^8 q^2}{429 l^{10}}\\
 && H^{(4,\pm4)}=-\frac{12288 \sqrt{\frac{10}{7}} \pi ^{3/2} d^8 q^2
   \left(r^2-l^2\right)}{143 l^{12}}, \quad K^{(4,\pm4)}= \frac{16384 \sqrt{\frac{10}{7}} \pi ^{3/2} d^8 q^2}{429
   l^{10}}
\end{eqnarray}
To find the horizon, we solve $g_{tt}=g_{tt}^{dS}+h_{tt}^{(0,0)}+h_{tt}^{(4,0)}+h_{tt}^{(4,4)}+h_{tt}^{(4,-4)}=0$. As before, only $L=0$ mode can change the horizon's area. The higher modes can only change the shape of the horizon in a manner that preserves the area. We take the ansatz for the horizon as $r_h'=l- 2m- \epsilon(\theta,\phi)$. For this configuration of charges and masses, the location and shape of the horizon (and plotted in fig. \ref{fig:molecule plot}) is
 \begin{eqnarray}
 \label{molecule shape}
    && r_h'=l-2m- \Bigg(\frac{8 m^2}{l}+\frac{64 m^3}{l^2}+\frac{512 m^4}{l^3}+\frac{4352 m^5}{l^4}+\frac{37888 m^6}{l^5}\nonumber\\
     &&-\frac{3 m^3 \left(5 d^4 \left(8 \sin ^4(\theta ) \cos (4
   \phi )+4 \cos (2 \theta )+7 \cos (4 \theta )\right)+9
   \left(d^4-12288 m^4\right)\right)}{l^6}\nonumber\\
   &&-\frac{4 m^4 \left(95 d^4 \left(8 \sin ^4(\theta ) \cos (4
   \phi )+4 \cos (2 \theta )+7 \cos (4 \theta )\right)+171
   d^4-727040 m^4\right)}{l^7}\nonumber\\
   &&-\frac{4 m^5 \left(1635 d^4 \left(8 \sin ^4(\theta ) \cos
   (4 \phi )+4 \cos (2 \theta )+7 \cos (4 \theta
   )\right)+2943 d^4-6373376 m^4\right)}{l^8}\nonumber\\
   &&+\frac{-\frac{4352}{189} \pi  d^8 q^2-94960 d^4 m^6 \left(8
   \sin ^4(\theta ) \cos (4 \phi )+4 \cos (2 \theta )+7
   \cos (4 \theta )\right)-170928 d^4 m^6+223477760
   m^{10}}{l^9}\Bigg)\nonumber\\
 \end{eqnarray}
 The electric field through the horizon is
 \begin{eqnarray}
    F^{tr}=\frac{d^4 q \left(40 \sin ^4(\theta ) \cos (4 \phi )+20 \cos (2 \theta )+35 \cos (4 \theta )+9\right)}{2 l^6}+O(1/l^7). 
    \label{eqn: mol field}
 \end{eqnarray}
  \textbf{The area and flux from the cap region:} The dip and protrusions on the horizon are defined relative to the area of the $L=0$ mode. We find the cap near the north pole is spanned by the region $\phi \in [0,2 \pi], \, \theta \in [0, \theta_{crit}]$, where $ \theta_{crit}\sim \pi/6$. Now we calculate the difference in the area of the top cap region relative to the area due to the monopole term. 
  \begin{eqnarray}
      \Delta A\sim\frac{27 \sqrt{3} \pi  d^4 m^3 (l-2 m)}{l^6}\geq 0
  \end{eqnarray}
To see the effect of charges, we need to go to $O(1/l^8)$, but these higher-order terms won't change the conclusion. The charges can be calculated using the Komar formula \eqref{komar}
 \begin{eqnarray}
 \label{whole cap}
     Q=\int_0^{2 \pi}\int_{0}^{\theta_{crit}} \,d \theta \, d \phi \, \, r_h^{2}\, \mathrm{sin}\, \theta F^{tr}\sim \frac{9 \sqrt{3} \pi  d^4 q (l-2 m)^2}{4 l^6}
 \end{eqnarray}
To see the effects of individual faces, we integrate in the $\phi$ direction to a restricted value.
\begin{eqnarray}
     Q_{face1}=\int_{\pi/4}^{3 \pi/4 }\int_{0}^{\theta_{crit}} \,d \theta \, d \phi \, \, r_h^{2}\, \mathrm{sin}\, \theta F^{tr}\sim \frac{9 \sqrt{3} \pi  d^4 q (l-2 m)^2}{16 l^6}\nonumber\\
 \end{eqnarray}
 It is $1/4$ of the charges coming from the whole capped region \eqref{whole cap}.\\
 
 The effects of the central charge $-8 q$ can be seen when we increase the value of $\theta_{cri}= \pi/4$.
 \begin{eqnarray}
  Q=  \frac{\sqrt{2} \pi  d^4 q (l-2 m)^2}{l^6}
 \end{eqnarray}
 The effective charge reduces as compared to \eqref{whole cap}. After setting the $\theta_{cri}= \pi/3$, the total charges coming from the entire region $\phi \in (0,2 \pi)$ is negative.
 \begin{eqnarray}
     Q= -\frac{15 \pi  d^4 q (l-2 m)^2}{4 l^6}
 \end{eqnarray}
 The contribution of individual side faces is $1/4$ of the above value. The contribution of the entire northern hemisphere is 0. A similar conclusion can be drawn for the southern hemisphere.

\section{Rotating objects}
\label{rotating formalism}

We now turn our attention to studying how rotation affects the cosmological horizon. Rotation is purely an effect of general relativity with no Newtonian counterpart, so we look to the Kerr-de Sitter solution studied in \cite{Akcay:2010vt} for consistency and guidance. A rotating object with a small angular momentum parameter $a\equiv J/M$ satisfying $a << M<<l $ in de Sitter spacetime can again be treated as perturbations about empty de Sitter spacetime. In the following analyses, we analyze vacuum solutions. To account for rotation, we need to use the axial perturbations as defined in \cite{Regge1957} and written in a more general form in section \ref{charged formalism}. In the Regge-Wheeler gauge, the axial components of the perturbed metric are given by 
\begin{eqnarray}
&&h_{t\theta}= -h_0(r) \frac{1}{\mathrm{sin}\, \theta} \,\, \partial_{\phi} Y_{LM}, \quad    h_{r\theta}= -h_1(r) \frac{1}{\mathrm{sin}\, \theta} \,\, \partial_{\phi} Y_{LM}\\
&&h_{t\phi}= h_0(r) \mathrm{sin}\, \theta \,\, \partial_{\theta} Y_{LM}, \quad    h_{r\phi}= h_1(r) \mathrm{sin}\, \theta \,\, \partial_{\theta} Y_{LM},
\end{eqnarray}
With the other components vanishing as long as we choose the rotations to be aligned with the $z$ axis. We will first show that a rotating object in de Sitter spacetime with angular parameter $a<<m<<l$ produces axial perturbation $h_{t\phi}$ which can be thought of as the slow rotation limit of Kerr-de Sitter black hole. Taking superpositions of single rotating objects, we build configurations of a rotating dipole and a cube (both have the total angular momentum zero) whose effect on the cosmological horizon we study.  

\subsection{A single rotating mass}

We first study an object with mass $m$ and rotating along the $z$ axis, placed at the origin of the static patch. This configuration is described by ($L=1$, $M=0$) mode perturbations. Solving the source-free Einstein's equations to linear order for only the axial perturbations, we find 
\begin{eqnarray}
    h_0(r)= \frac{c_1}{r}+r^2 c_2, \quad  h_1(r)=0.
\end{eqnarray}
To fix the constants of integration, we match our solution to the slow-rotation limit of Kerr-de Sitte,r which has metric component $g_{t\phi}$ as
\begin{eqnarray}
 g_{t\phi}^{Kerr}= -\frac{a \sin ^2(\theta ) \left(2 l^2
   m+r^3\right)}{l^2 r}.
\end{eqnarray}
Then constants of integration are fixed to be
\begin{eqnarray}
    c_1= 4\, a\,  m\,  \sqrt{\frac{\pi}{3}}, \quad c_2 =\frac{2 a \sqrt{\pi}}{l^2 \sqrt{3}}.
\end{eqnarray}
To describe the rotating mass we must also use the polar perturbations, which in this case will describe the Schwarzschild-de Sitter solution by Birkhoff's theorem. The details of this calculation can be found by taking $q\to 0$ in the Reissner-Nordstr\"{o}m-de Sitter solution in section \ref{charged objects} or found in \cite{Fischler:2024cgm}. As before, we need to redefine $r'=r \sqrt{\frac{2m}{r}+1} \sim r+m$ to match with Schwarzschild-de Sitter radial coordinate. 
After superimposing the polar perturbation, we find the metric component $g_{tt}$ has zeros at $r'_{h}=l-m$. Hence, the horizon is located at $r'_{h}=l-m$. We however find a non-trivial $g_{t\phi}$ component of the metric, 
\begin{eqnarray}
\label{l1gtphi}
    g_{t\phi}=-\frac{a \sin ^2(\theta ) \left(2 l^2 m+(r'-m)^3\right)}{l^2
   (r'-m)}.
\end{eqnarray}

\textbf{Angular momentum charge at the horizon:-} We find the angular momentum charge for this perturbed spacetime \eqref{l1gtphi} using the Komar formula \cite{Carroll:2004st,Misner:1973prb}.
\begin{eqnarray}
    J=-\frac{1}{8 \pi G} \int_{\partial \Sigma} d^2 x \sqrt{\gamma^{(2)}} n_{\mu} \sigma_{\nu} \nabla^{\mu} R^{\nu},
\end{eqnarray}
where $R=\partial_{\phi}$ is the rotational Killing vector. The normal vector $n_{\mu}$ and normal to the boundary $\sigma_{\mu}$ is calculated for empty de Sitter spacetime (see the discussion near \eqref{komar} for more clarifications for using this formula). 
 \begin{eqnarray}
     &&n_0=-(1-\frac{r^2}{l^2})^{\frac{1}{2}},\quad  n_i=0, \quad n_{\mu}n^{\mu}=-1\nonumber\\
     &&\sigma_1= (1-\frac{r^2}{l^2})^{\frac{-1}{2}}, \quad \sigma_0=\sigma_2=\sigma_3=0,\quad \sigma_{\mu}\sigma^{\mu}=1
 \end{eqnarray}
 Hence, the conserved angular momentum charge for the metric with additional rotation component \eqref{l1gtphi} is found to be
 \begin{eqnarray}
     J=- a m.
 \end{eqnarray}
We see that the cosmological horizon counter-rotates to the object deep within the de Sitter bulk. 
\subsection{A dipole}
We now repeat the above exercise with two rotating objects at the end of the dipole (with separation $d$ and mass $m$) but in opposite directions. This is the analog of the dipole case studied earlier, but now with rotation rather than charge. The polar perturbations that describe this mass configuration can again be seen by taking $q\to 0$ in the charged dipole example or corresponding to the massive binary studied in \cite{Fischler:2024cgm}. This spacetime has been studied numerically in \cite{dias2024spinningblackbinariessitter}. The linearized equation of motion for the perturbations is
\begin{eqnarray}
    r^2 (l-r) (l+r) h_0''(r)+2 h_0(r) \left(r^2-3
   l^2\right)=0
\end{eqnarray}
The equation of motion for $h_1(r)$ is trivial, and it can be set to zero. The solution $h_0(r)$ can be found, and the constant of integration can be matched with the effective axial perturbation due to the rotating dipole. The effective axial perturbation is obtained by superimposing the two individual rotating objects, but rotating in opposite directions. The perturbation due to individual objects is just the slow rotation limit of Kerr-de Sitter.
\begin{eqnarray}
 g_{t\phi}= -\frac{a \sin ^2(\theta ) \left(2 l^2
   m+r^3\right)}{l^2 r}.
\end{eqnarray}
For a dipole, we superpose the above solution separated by a distance $d$, and rotation in opposite directions, which yields the metric component in the multipole regime as
\begin{eqnarray}
  \delta g_{t\phi}^{dipole}= \frac{a d \sin ^2(\theta ) \cos (\theta ) \left(2 r^3-l^2
   m\right)}{l^2 r^2}.
\end{eqnarray}
The above potential fixes the constants of integration as
\begin{eqnarray}
    c_1= -\frac{2 \sqrt{\frac{\pi }{5}} a d m}{3 l^2}, \quad c_2=0
\end{eqnarray}
Then the solution for the Regge-Wheeler equation can be written as
\begin{eqnarray}
  h_0(r)=-\frac{2 \sqrt{\frac{\pi }{5}} a d m \left(-3 l^4+2 l^2 r^2+r^4\right)}{9 l^4 r^2} 
\end{eqnarray}
It yields the total metric perturbation due to the rotating dipole as 
\begin{eqnarray}
    h_{t\phi}=\frac{a d m \sin ^2(\theta ) \cos (\theta ) \left(-3 l^4+2 l^2 r^2+r^4\right)}{3 l^4 r^2}
\end{eqnarray}
The shape and location of the horizon are already found in eq \eqref{horizon dipole} by setting $q=0$ (see fig. \ref{fig:dipole plot} for the shape of the horizon). Here, due to the angular momentum, we have a non-trivial $g_{t\phi}$ component on the horizon. Now, we calculate the angular momentum charges on the horizon.\\

\textbf{Angular momentum from the dip region: } The total angular momentum from the dip region (near the north pole- see fig. \ref{fig:dipole plot}). 
can be calculated using the Komar formula but now $\theta \in [0,\theta_{crit}]$, where 
$\theta_{crit}\sim \frac{\pi}{4}$.
\begin{eqnarray}
     J\sim -\Bigg(\frac{a \,d\, m}{48 l}+\frac{ 
  a d  m^2}{12 l^2}\Bigg)+O(\frac{1}{l^3}).
 \end{eqnarray}
 The integral  over the entire northern hemisphere produces 
 \begin{eqnarray}
     J\sim -\Bigg(\frac{a \,d\, m}{12 l}+\frac{ 
  a d  m^2}{3 l^2}\Bigg)+O(\frac{1}{l^3}).
 \end{eqnarray}
The total angular momentum over the entire horizon is zero.

\subsection{A cube}
Now we discuss the rotating cube. We place 4 objects with mass $m/8$ and $J= a m/8 \, \hat{k}$. And on the bottom face, we place 4 objects with mass $m/8$ and $J= -a m/8 \, \hat{k}$. Now the top and bottom faces have net angular momentum \footnote{This configuration is different than the cube case of fig. \ref{fig:cube config}. Here, on the top face, all 4 particles rotate in the same direction, while on the bottom face, all 4 rotate in opposite directions. The total net angular momentum of this cube is zero. One can discuss the rotating cube of fig. \ref{fig:cube config} but it is more tedious than the case discussed in this section.}. The effective axial perturbation is obtained by superimposing the individual rotating objects. 
\begin{eqnarray}
\label{cube_rot}
    h^{cube}_{t\phi}=\frac{a \,d\, m \sin ^2(\theta ) \cos (\theta ) \left(5 d^2 \cos
   (2 \theta )-d^2-2 r^2\right)}{r^4}
\end{eqnarray}
The Einstein equation gives the following differential equation for the perturbation in $L=4$ mode
\begin{eqnarray}
    r^2 (l-r) (l+r) h_0''(r)+2 h_0(r) \left(r^2-10
   l^2\right)=0.
\end{eqnarray}
The solution of the above differential equation is 
\begin{eqnarray}
  h_0(r)=  \frac{i c_2 r^5 \, _2F_1\left(\frac{3}{2},3;\frac{11}{2};\frac{r^2}{l^2}\right)}{l^5}+\frac{c_1 \left(35 l^6-45 l^4 r^2+9 l^2 r^4+r^6\right)}{35
   l^2 r^4}.
\end{eqnarray}
The constants of integration can be matched with the metric perturbation due to a rotating cube \eqref{cube_rot}.
\begin{eqnarray}
   c_1= \frac{8 \sqrt{\pi } a d^3 m}{15 l^4},\quad  c_2=0
\end{eqnarray}
Hence, the metric perturbation due to the rotating cube is 
\begin{eqnarray}
 h_{t\phi}= h_0(r)\mathrm{sin}\, \theta \partial_{\theta} Y_{4,0}=-\Bigg(\frac{ \left(35 l^6-45 l^4 r^2+9 l^2 r^4+r^6\right)}{35
   l^2 r^4}  \frac{8 \sqrt{\pi } a d^3 m}{15 l^4}\Bigg)\frac{15 \sin ^2(\theta ) \cos (\theta ) (7 \cos (2 \theta )+1)}{8 \sqrt{\pi }}\nonumber\\
\end{eqnarray}

The location and shape of the horizon can be obtained by setting $q=0$ in the \eqref{horizon cube} and see the fig. \ref{fig:cube diff} for horizon shape. We can find the contributions to angular momentum at the protrusion (cap near the north pole) region. Here the $\theta \in [0,\theta_{crit}], \,\text{and}\,\, \theta_{crit} \sim \frac{\pi}{6}$. Using the Komar formula, we can find the total angular momentum from the cap region as
\begin{eqnarray}
    J\sim -\Bigg(\frac{17 a d^3 m}{1120 l^3}+\frac{153 a d^3 m^2}{560 l^4}\Bigg)+O(\frac{1}{l^5})
\end{eqnarray}
The contributions from the individual constituents are $1/4$ of the above value. When one extends the range of $\theta_{crit} \sim \pi/2$ to include the contributions of other dips and  protrusions, then the angular momenta are
\begin{eqnarray}
    J\sim \Bigg(\frac{2 a d^3 m}{35 l^3}+\frac{36 a d^3 m^2}{35 l^4}\Bigg)+O(\frac{1}{l^5})
\end{eqnarray}
Again, when we integrate over the whole sphere, the net angular momentum is zero.


\section{Discussion}
In this paper, we have shown how small charges and rotation ($a=J/M$ small) influence the de Sitter cosmological horizon. Using the Regge-Wheeler formalism, we expanded upon the work in \cite{Fischler:2024cgm} and calculated corrections to the location and shape of the cosmological horizon for configurations of charged and rotating masses in static albeit unstable equilibrium within the de Sitter bulk. The horizon remains dual to the solid within the bulk, but deformations of the cosmological horizon are also sensitive to the energy stored in the electric fields of the bulk configurations (as is well known in the case of Reissner-Nordstr\"{o}m-de Sitter). Since the charged matter configurations must be invariant under the same symmetry group as the masses within the bulk, the electric field energy can only enhance the effect of the masses on the cosmological horizon, but not change its shape.  On the other hand, the horizon location does not depend on the rotation of the masses (this statement is only valid to leading order in the slow rotation limit of Kerr-de Sitter).  

Although we analyzed neutral charge configurations and rotation configurations with a net angular momentum of 0, we have shown that the cosmological horizon inherits an induced charge or angular momentum polarization to the bulk configuration. From such horizon data, we can determine differences in the charge configurations within the bulk as can be seen from the electric fields at the horizon of the cube with alternating charges \ref{eqn: cube fields} and the `crystalline atom' configuration \ref{eqn: mol field}. The field lines start from the charges in the bulk and end on the cosmological horizon. These field lines induce charge (smeared) on the horizon consistent with the Gauss law. More generally, measurements of the cosmological horizon are indeed sensitive to all the details of classical black holes within the de Sitter bulk allowed by the no-hair theorem. 

In this work, we only address static configurations placed near the center of the static patch. While the configurations are unstable, important lessons can be gleaned from the behavior of the cosmological horizon, as was done for the similarly unstable Schwarzschild-de Sitter solution. It would be instructive to see how dynamics within the bulk of de Sitter affect the ``horizon". Studying dynamic configurations, such as an orbiting binary, could be of interest due to their increased stability. Moreover, a shockwave in de Sitter is an exact solution of Einstein's equations \cite{Hotta:1992qy}, and understanding the behavior of the ``cosmological horizon" for such spacetimes is paramount to the development of a quantum theory of de Sitter spacetime. 

Our results should be interpreted as providing (semi-)classical data for a holographic description of de Sitter spacetime. We have shown that the cosmological horizon contains
more information than just the entropy of the bulk and can be used to fully determine (classically) the matter present within the de Sitter bulk. However, we should be concerned with the holographic description involving quantum degrees of freedom living on the horizon and how they might be able to reproduce these results.  
\label{discussions}

\begin{acknowledgments}
The work of WF, HK and SR are supported by NSF grant PHY-2210562. During the preparation of this manuscript, SR began a position at Sarah Lawrence College which she thanks for additional support.
\end{acknowledgments}


\appendix
\section{Decomposition of stress tensor into tensor harmonics}
\label{stress_decom}
The decomposition of a symmetric second-rank tensor into tensor harmonics can be written as (conventions; for more details, see Zerilli's \cite{Zerilli:1970wzz}).
\begin{eqnarray}
\label{decomposition_stress}
T_{ab}=\sum_{L,M}&& A^{(0)}_{LM} a^{(0)}_{LM}+A^{(1)}_{LM} a^{(1)}_{LM}+A_{LM} a_{LM}+B^{(0)}_{LM}b^{(0)}_{LM}+B_{LM} b_{LM}\nonumber\\
&&+Q^{(0)}_{LM} c^{(0)}_{LM}+Q_{LM} c_{LM}+G_{LM} g_{LM}+D_{LM} d_{LM}+F_{LM} f_{LM}\nonumber\\
\end{eqnarray}
In the above equation, the ``capital'' letters like $A_{LM}$ correspond to mode functions, and the ``small'' letters like $a_{LM}$ are irreducible second-rank tensors. These basis tensors can be written as

 $ a^{(0)}_{LM}=\left(\begin{array}{cccc}
 Y_{LM} & 0 & 0 & 0 \\
 0 & 0 & 0 & 0 \\
 0 & 0 & 0 & 0 \\
 0 & 0 & 0 & 0 \\
\end{array}
\right)  $,  $\quad  a_{LM}=\left(\begin{array}{cccc}
 0 & 0 & 0 & 0 \\
 0 & Y_{LM} & 0 & 0 \\
 0 & 0 & 0 & 0 \\
 0 & 0 & 0 & 0 \\
\end{array}
\right)  $,  \\

\vspace{ 1 cm}
$
 a^{(1)}_{LM}=\frac{i}{\sqrt{2}}\left(\begin{array}{cccc}
  0& Y_{LM} & 0 & 0 \\
 Y_{LM} & 0 & 0 & 0 \\
 0 & 0 & 0 & 0 \\
 0 & 0 & 0 & 0 \\
\end{array}
\right), \quad $ $ g_{LM}=\frac{r^2}{\sqrt{2}}\left(\begin{array}{cccc}
 0 & 0 & 0 & 0 \\
 0 & 0 & 0 & 0 \\
 0 & 0 & 1 & 0 \\
 0 & 0 & 0 & \mathrm{sin}^2 \theta \\
\end{array}
\right) Y_{LM}$\\

\vspace{ 1 cm}

$b^{(0)}_{LM}=\frac{i r}{\sqrt{2 L(L+1)}}\left(\begin{array}{cccc}
  0& 0 & \partial_{\theta}Y_{LM} & \partial_{\phi}Y_{LM} \\
 0 & 0 & 0 & 0 \\
 * & 0 & 0 & 0 \\
 * & 0 & 0 & 0 \\
\end{array}
\right), \quad $ $b_{LM}=\frac{ r}{\sqrt{2 L(L+1)}}\left(\begin{array}{cccc}
  0& 0 & 0 &0  \\
 0 & 0 & \partial_{\theta}Y_{LM} & \partial_{\phi}Y_{LM} \\
 0 & * & 0 & 0 \\
 0 & * & 0 & 0 \\
\end{array}
\right) $ \\

\vspace{ 1 cm}

$c^{(0)}_{LM}=\frac{ r}{\sqrt{2 L(L+1)}}\left(\begin{array}{cccc}
  0& 0 & \frac{1}{\mathrm{sin}\, \theta}\partial_{\phi}Y_{LM} & -\mathrm{sin}\, \theta \partial_{\theta}Y_{LM} \\
 0 & 0 & 0 & 0 \\
 * & 0 & 0 & 0 \\
 * & 0 & 0 & 0 \\
\end{array}
\right), $  \\

\vspace{1 cm}
$c_{LM}=\frac{ i r}{\sqrt{2 L(L+1)}}\left(\begin{array}{cccc}
  0& 0 & 0 &0  \\
 0 & 0 & \frac{1}{\mathrm{sin}\, \theta}\partial_{\phi}Y_{LM} & -\mathrm{sin}\, \theta \partial_{\theta}Y_{LM} \\
 0 & * & 0 & 0 \\
 0 & * & 0 & 0 \\
\end{array}
\right) $\\

\vspace{1 cm}
$  d_{LM}=\frac{-i r^2}{[2 L(L+1)(L-1)(L+2)]^{\frac{1}{2}}}\left(\begin{array}{cccc}
 0 & 0 & 0 & 0 \\
 0 & 0 & 0 & 0 \\
 0 & 0 &-\frac{1}{\mathrm{sin}\, \theta} X_{LM} & \mathrm{sin}\, \theta W_{LM} \\
 0 & 0 &*& \mathrm{sin}\, \theta  X_{LM} \\
\end{array}
\right)  $\\

\vspace{1 cm}

$  f_{LM}=\frac{r^2}{[2 L(L+1)(L-1)(L+2)]^{\frac{1}{2}}}\left(\begin{array}{cccc}
 0 & 0 & 0 & 0 \\
 0 & 0 & 0 & 0 \\
 0 & 0 & W_{LM} & X_{LM} \\
 0 & 0 & X_{LM}& \mathrm{sin}^2 \theta\,  W_{LM} \\
\end{array}
\right)  $

Here \begin{eqnarray}
    X_{LM}=2 \frac{\partial}{\partial \phi}(\frac{\partial}{\partial \theta}- \mathrm{cot} \theta) Y_{LM}, \quad   W_{LM}= (\frac{\partial^2}{\partial \theta^2}-\mathrm{cot} \theta \frac{\partial}{\partial \theta} -\frac{1}{\mathrm{sin}^2 \theta} \frac{\partial^2}{\partial \phi^2}) Y_{LM}
\end{eqnarray}
These basis functions are orthonormal and complete. The inner product of these functions can be written as
\begin{eqnarray}
    \langle T, S \rangle \equiv \int T^{*}:S\,  d \Omega, \quad T:S \equiv g^{\mu\alpha} g^{\nu\beta} T_{\mu\nu} S_{\alpha\beta}
\end{eqnarray}
Hence the mode functions can be written as
\begin{eqnarray}
    A_{LM}= \langle a_{LM}, T\rangle, \quad  G_{LM}= \langle g_{LM}, T \rangle
\end{eqnarray}
Zerilli \cite{Zerilli:1970wzz} has provided a map between Regge-Wheeler \cite{Regge1957} gauge and his basis. One can gauge fix and write the perturbation in the Regge-Wheeler gauge.

\bibliographystyle{JHEP}

\bibliography{ref}
\end{document}